# Real-time CBCT Imaging and Motion Tracking via a Single Arbitrarily-angled X-ray Projection by a Joint Dynamic Reconstruction and Motion Estimation (DREME) Framework

Hua-Chieh Shao[1], Tielige Mengke[1], Tinsu Pan[2], You Zhang[1]

[1]The Medical Artificial Intelligence and Automation (MAIA) Laboratory
Department of Radiation Oncology, University of Texas Southwestern Medical Center, Dallas, TX 75390, USA
[2]Department of Imaging Physics
University of Texas MD Anderson Cancer Center, Houston, TX, 77030, USA

Corresponding address:

You Zhang
Department of Radiation Oncology
University of Texas Southwestern Medical Center
2280 Inwood Road
Dallas, TX 75390
Email: You.Zhang@UTSouthwestern.edu
Tel: (214) 645-2699

**Abstract**

**Objective:** Real-time cone-beam computed tomography (CBCT) provides instantaneous visualization of patient anatomy for image guidance, motion tracking, and online treatment adaptation in radiotherapy. While many real-time imaging and motion tracking methods leveraged patient-specific prior information to alleviate under-sampling challenges and meet the temporal constraint (< 500 ms), the prior information can be outdated and introduce biases, thus compromising the imaging and motion tracking accuracy. To address this challenge, we developed a framework (DREME) for real-time CBCT imaging and motion estimation, without relying on patient-specific prior knowledge.

**Approach:** DREME incorporates a deep learning-based real-time CBCT imaging and motion estimation method into a dynamic CBCT reconstruction framework. The reconstruction framework reconstructs a dynamic sequence of CBCTs in a data-driven manner from a standard pre-treatment scan, without utilizing patient-specific knowledge. Meanwhile, a convolutional neural network-based motion encoder is jointly trained during the reconstruction to learn motion-related features relevant for real-time motion estimation, based on a single arbitrarily-angled x-ray projection. DREME was tested on digital phantom simulation and real patient studies.

**Main results:** DREME accurately solved 3D respiration-induced anatomic motion in real time (~1.5 ms inference time for each x-ray projection). In the digital phantom study, it achieved an average lung tumor center-of-mass localization error of 1.2±0.9 mm (Mean±SD). In the patient study, it achieved a real-time tumor localization accuracy of 1.8±1.6 mm in the projection domain.

**Significance:** DREME achieves CBCT and volumetric motion estimation in real time from a single x-ray projection at arbitrary angles, paving the way for future clinical applications in intra-fractional motion management. In addition, it can be used for dose tracking and treatment assessment, when combined with real-time dose calculation.

## 1. Introduction

Radiotherapy aims to maximize local tumor control while minimizing radiotoxicity to the healthy tissues/organs adjacent to target tumors (Verellen et al. 2007). To accomplish such goals, sophisticated treatment planning and dose delivery techniques, such as volumetric modulated arc therapy (Rashid et al. 2021) and stereotactic body radiotherapy (Kimura et al. 2022), were developed to yield dose distributions highly conformal to target tumors (Bernier et al. 2004). However, respiration-induced anatomic motion, particularly for tumors in the thoracic and abdominal regions, introduces variations in target shapes and locations (Shirato et al. 2004), compromising the precision of radiotherapy. To address the motion-related uncertainties, patient-specific motion characteristics (e.g., motion patterns and amplitudes) are often quantified during treatment simulation and planning, based on which customized motion management strategies are applied to patients during the treatment (Keall et al. 2006). Most motion management tools (e.g., respiratory gating, deep inspiration breath hold, or tumor tracking) require real-time imaging/motion signals to monitor daily patient motion when delivering each treatment (Keall et al. 2019). However, until today, most of the monitoring signals are limited to surrogate or two-dimensional (2D) signals, including surface optical markers, interstitial fiducial markers, or x-ray fluoroscopy (Cui et al. 2007, Xu et al. 2014, Poulsen et al. 2015, Sakata et al. 2020). These 1D or 2D signals cannot not accurately reflect the 3D deformable motion (Roman et al. 2012), or capture complicated nonlinear motion trajectories of organs and tumors (Langen and Jones 2001, Seppenwoolde et al. 2002, Shirato et al. 2004). Therefore, 3D volumetric imaging is highly desired to capture instantaneous patient anatomy to achieve the most accurate tumor localization, allow intra-treatment dose tracking, and enable potential real-time treatment adaptation.

Currently, a major limitation of x-ray-based volumetric imaging is its temporal resolution. For respiratory motion (typically 3-5 seconds per cycle), the AAPM task group 264 recommended real-time tumor tracking or plan adaptation to have a temporal resolution of ≤500 milliseconds (ms) (Keall et al. 2021). Within such a timeframe, very few 2D x-ray projections can be acquired. Reconstructing volumetric CBCT images from these projections is not feasible by conventional reconstruction algorithms due to extreme under-sampling. Recently, the interest is growing in using deep leaning (DL)-based approaches to solve this problem (Mylonas et al. 2021, Liu et al. 2024). With usually-low inference latency, DL solutions are seen particularly suited for the real-time imaging problem. One type of DL approaches (Type I) attempts direct 3D image reconstruction from sparsely sampled 2D x-ray projections (Shen et al. 2019, Ying et al. 2019, Tong et al. 2020, Shen et al. 2022, Zhou et al. 2022, Zhang et al. 2024, Zhu et al. 2024). In particular, Shen et al. (Shen et al. 2019) developed a patient-specific volumetric reconstruction method from a single x-ray projection. Their network used a representation module to extract image features from a 2D x-ray projection, and then converting the 2D feature maps by a transformation module and a generation module to 3D feature maps and synthesizing the 3D image. Ying et al. (Ying et al. 2019) proposed to reconstruct a CT volume from two orthogonal x-ray projections, using generative adversarial learning. Their network comprised two 2D encoders and a 3D decoder that were linked by skip connections to bridge the 2D and 3D feature maps and to fuse the features extracted from the two orthogonal projections. However, reconstructing a 3D volume from a single or few 2D projections is a highly ill-posed inverse problem, thus the DL models can suffer from instability and generalizability issues. Moreover, the aforementioned networks were trained on x-ray projections at fixed gantry angles, rendering them impractical for radiotherapy treatments with rotating gantries.

The second type of DL works (Type II) utilizes deformable registration for real-time image/motion estimation and target localization (Wei et al. 2020, Nakao et al. 2022, Shao et al. 2023). The registration-based approaches introduce prior knowledge (e.g., known patient-specific image and/or motion model), and infer 3D motion fields from limited on-board signals (e.g., one x-ray projection) to deform the patient-specific reference image to generate new images. The prior knowledge supplies additional (patient-specific) information to condition the ill-posed problem, and allows target tumors to be directly localized after registration through propagations of prior segmentations (thus removing the need of additional segmentations). In particular, Wei et al. (Wei et al. 2020) proposed a patient-specific CNN to solve 3D motion fields from a single x-ray projection at arbitrary gantry angles. The angle-agnostic imaging was achieved by using an angle-dependent lung mask to extracted motion-related image features, followed by angle-dependent fully connected layers to map 2D image features to 3D motion fields. However, training fully connected layers separately for each projection angle is less practical and highly resource demanding. Recently, Shao et al. (Shao et al. 2023) developed a graph neural network-based, angle-agnostic DL model to estimate 3D deformable motion from an arbitrarily-angled x-ray projection. Instead of using angle-specific model layers, they incorporated the x-ray cone-beam projection geometry into the feature extraction process to allow patient-specific, angle-agnostic inference. However, although the registration-driven real-time imaging solutions hold great promises, they similarly face generalizability and robustness challenges. Generally, DL models, either type I or type II, require a large training dataset to properly learn and generalize to unseen scenarios. Due to the ill-condition of real-time imaging, most of such models are developed as patient-specific models to more effectively capture the patient traits. For patient-specific models, to overcome the size limits of patient-specific data, they usually rely on prior motion models to augment the training samples (e.g., (Shen et al. 2019, Wei et al. 2020, Shao et al. 2023)). Typically, these motion models were derived from 4D-CTs acquired during treatment planning, and the 4D-CTs were augmented with the motion model to yield more random motion states. The motion model extracted from a planning 4D-CT, however, may not reflect varying breathing motion patterns observed during different treatments (Vergalasova and Cai 2020). The 4D-CT motion model also only captures breathing-related intra-scan motion, while other day-to-day anatomical motion cannot be modeled. The deformation-driven augmentations cannot model non-deformation-related anatomical variations, either (Zhang et al. 2017). In addition, 4D-CT is commonly acquired through retrospective phase sorting, under the assumption of regular and periodic motion. However, patient motion can be highly irregular, leading to residual 4D-CT motion artifacts that further degrade the accuracy of its motion model (Yasue et al. 2022).

To overcome the limitations and challenges of previous methods, in this study we proposed a DL-based joint framework, dynamic reconstruction and motion estimation (DREME), for real-time CBCT imaging and tumor localization. DREME integrates a DL-based motion estimation module into a previously-developed dynamic CBCT reconstruction workflow (Shao et al. 2024), solving two learning-based tasks in a single training session: 1. reconstructing a pre-treatment, dynamic CBCT sequence to capture daily anatomy and motion patterns; and 2. deriving a CNN motion encoder to allow real-time CBCT and motion inference from an arbitrarily-angled, intra-treatment x-ray projection. The first learning task reconstructs a dynamic sequence of CBCT volumes (i.e., one CBCT volume for each x-ray projection) from a standard pre-treatment cone-beam scan to extract the most up-to-date patient anatomic and motion information without relying on patient-specific prior knowledge (Shao et al. 2024). The second learning task uses the motion model directly learned from the pre-treatment dynamic CBCT reconstruction to derive a light-weight CNN motion encoder for real-time motion estimation, which can be used to deform the dynamic CBCT to real-time CBCT images. Compared with the previous DL-based real-time imaging methods, the dual task learning-based DREME framework includes both reconstruction and registration components, allowing it to combine the benefits from both worlds. Compared with the type I DL-based reconstruction methods, DREME deforms the real-time CBCT from a reconstruction-based pre-treatment CBCT scan, by solving motion from real-time singular x-ray projections. The pre-treatment CBCT brings in the most relevant prior information (immediate images of the same treatment session) to strongly condition the real-time imaging problem and stabilize its solution, without introducing generalizability issues. Under such conditioning, we are able to develop an angle-agnostic model, which has been challenging for previous type I methods. Compared with the type II DL-based registration methods, DREME uses newly reconstructed, pre-treatment dynamic CBCTs rather than planning 4D-CT images for motion modeling. The motion model learned from daily pre-treatment scans is not susceptible to day-to-day motion pattern variations. As the motion model is built on the pre-treatment dynamic CBCT images, any day-to-day anatomic variations, either deformation- or non-deformation-related, are naturally absorbed into the pre-treatment CBCTs and require no additional modeling. In addition, compared with 4D-CT, the dynamic CBCT of DREME is reconstructed to resolve both regular and irregular motion (Zhang et al. 2023, Shao et al. 2024), thus their images and the corresponding motion models are less susceptible to artifacts caused by irregular motion.

To the best of our knowledge, DREME is the first dual task learning framework capable of real-time CBCT imaging and motion estimation without relying on patient-specific prior knowledge for model training. It makes a 'one-shot' learning technique, which only requires a single pre-treatment CBCT scan for training and does not need additional projections/images/motion models, rendering it highly data-efficient. We evaluated the DREME framework using a digital phantom simulation study and a multi-institutional lung patient dataset. Ablation studies were also performed on DREME to assess the contribution of its individual components, along with comparisons with other real-time imaging solutions.

## 2. Materials and Methods

2.1 Overview of the dual task learning DREME framework

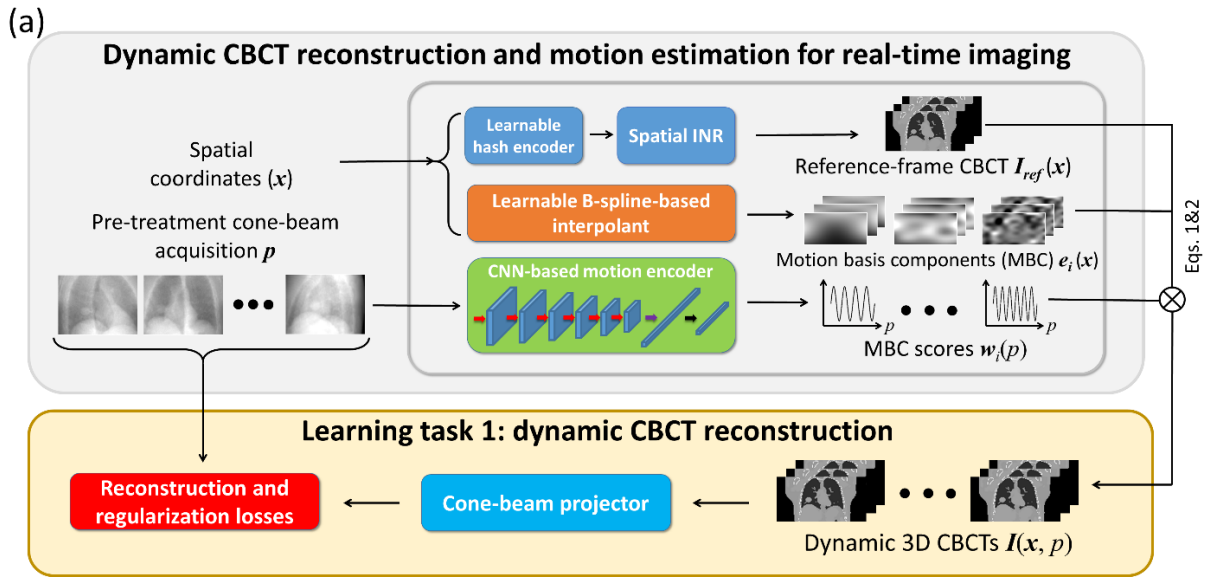

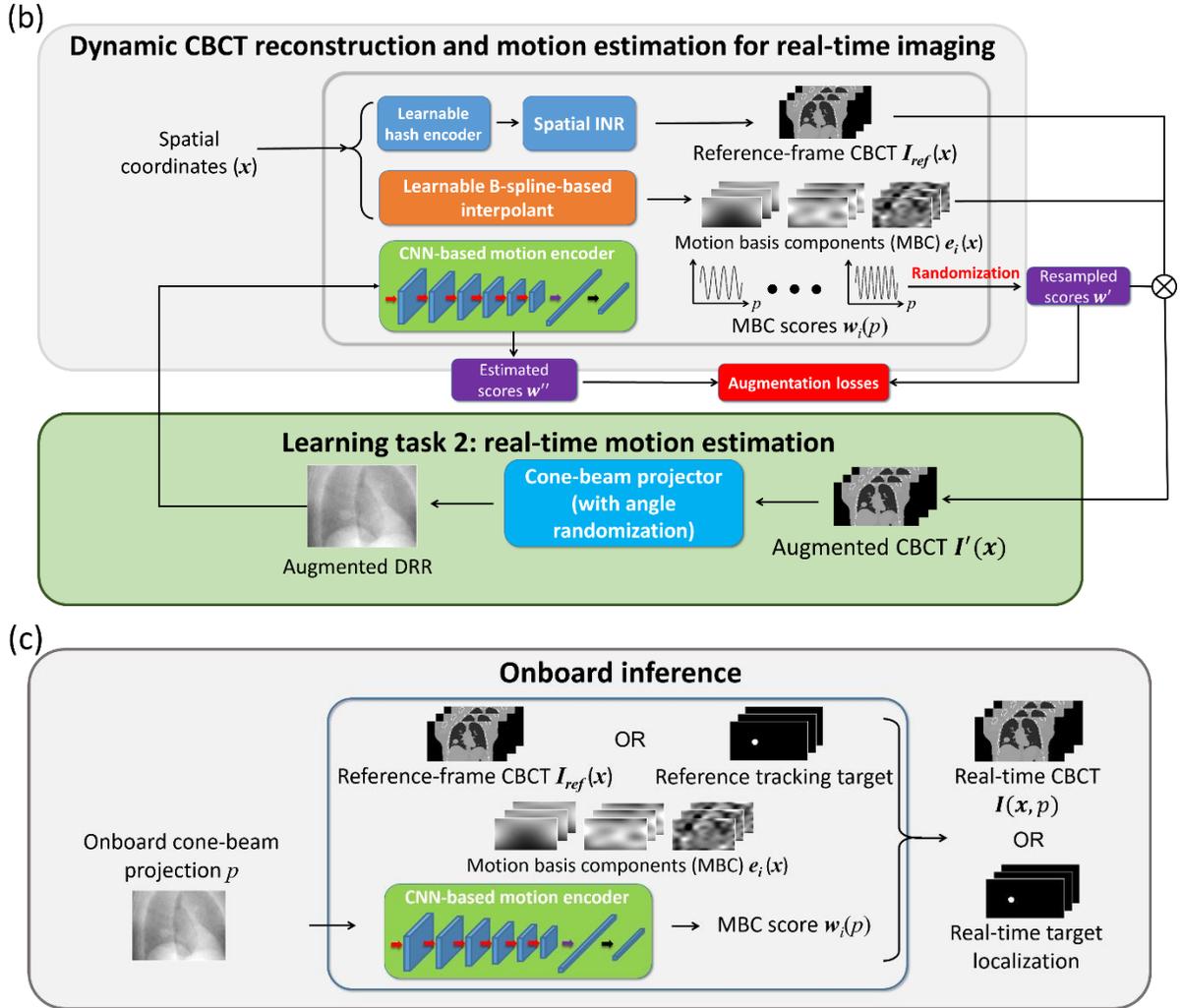

**Figure 1**. The DREME network architecture for dynamic CBCT reconstruction and real-time motion estimation. **(a)** DREME adopts a motion-compensated CBCT reconstruction approach based on a standard pre-treatment cone-beam acquisition $\boldsymbol{p}$, by joint deformable registration and reconstruction. The spatial implicit neural representation (INR) estimates the reference-frame CBCT $I_{ref}(x)$, and the CNN-based motion encoder, together with the learnable B-spline interpolant, estimates the deformation vector fields (DVFs) with respect to the reference CBCT. The dynamic CBCTs $I(x, p)$ are derived by deforming the reference CBCT using the solved DVFs corresponding to each projection $p$ of the cone-beam acquisition $\boldsymbol{p}$. The learning task of dynamic CBCT reconstruction is driven by maximizing the similarity between the digitally reconstructed radiographs (DRRs) and the corresponding cone-beam projections. To regularize the ill-posed spatiotemporal inverse problem, regularization losses are incorporated into the training objectives. **(b)** For the learning task of real-time imaging, deformable motion augmentation simulates random motion states by resampling MBC scores $w_i(p)$ to enhance the model's robustness to unseen motion states. In addition, projection angle augmentation is implemented to simulate DRRs of motion-augmented CBCTs at random projection angles, to promote angle-agnostic learning. **(c)** During the onboard inference stage, the motion encoder takes an onboard projection at an arbitrary angle as input and outputs the MBC scores to derive the real-time CBCT or 3D target geometry, based on the patient anatomy $I_{ref}(x)$/tracking target and the motion model (MBCs).

DREME achieves real-time CBCT imaging and markerless tumor tracking by combining dynamic CBCT reconstruction and real-time motion estimation into a single framework (Fig. 1), without requiring any patient-specific prior information. The reconstruction task (learning task 1) solves the latest patient anatomy as well as a motion model immediately before each treatment, which are used by the real-time imaging task (task 2) for motion tracking and real-time CBCT rendering. The framework was adapted from our previous work of dynamic CBCT reconstruction (Shao et al. 2024), with new components incorporated to allow simultaneous image reconstruction and motion encoder learning.

In general, DREME reconstructs a dynamic sequence of 3D CBCT volumes from a pre-treatment cone-beam projection set $\boldsymbol{p}$. For each CBCT scan, the anatomy captured at different time points are highly correlated. Therefore, the reconstruction algorithm took a joint deformable registration and reconstruction approach, assuming that the dynamic anatomy can be represented by a static reference anatomy $\boldsymbol{I}_{ref}(\boldsymbol{x})$ and a time-varying, projection-dependent deformable vector field (DVF) $\boldsymbol{d}(\boldsymbol{x}, p)$ with respect to the reference anatomy, i.e.,

$$\boldsymbol{I}(\boldsymbol{x}, p) = \boldsymbol{I}_{ref}(\boldsymbol{x} + \boldsymbol{d}(\boldsymbol{x}, p)), \tag{1}$$

where $\boldsymbol{x}$ denotes the voxel coordinates. Since reconstructing a dynamic sequence of 3D volumes can involve solving more than $10^8$ unknowns, dimension reduction is used to condition the ill-posed spatiotemporal inverse problem and avoid sub-optimal solutions. Particularly, we decomposed each DVF into low-rank spatiotemporal components by a data-driven motion model learned on the fly:

$$\boldsymbol{d}(\boldsymbol{x}, p) = \sum_{i=1}^{3} \boldsymbol{w}_i(p) \times \boldsymbol{e}_i(\boldsymbol{x}), \tag{2}$$

where $\boldsymbol{w}_i(p)$ is represented by a CNN-based motion encoder (Fig. 1), while the spatial components $\boldsymbol{e}_i(\boldsymbol{x})$ is represented by B-spline-based interpolants. We used three levels ($i$ = 1, 2, 3) for the spatial decomposition, which has been demonstrated sufficient for representing respiratory motion (Li et al. 2011). The control points of the B-spline interpolants are learnable parameters (Tegunov and Cramer 2019), and a hierarchical multiresolution strategy (Shao et al. 2024) was used so that each level $i$ represents a different spatial scale. The number of the control points is doubled each higher level to represent finer motion. The spatial components serve a basis set spanning a Hilbert sub-space of the solution space, thus called motion basis components (MBCs) in this work. The motion encoder takes an x-ray projection $p$ as an input, and estimates the corresponding MBC scores (or temporal coefficients) $\boldsymbol{w}_i(p)$ to map motion. The motion encoder is trained to decipher the pre-treatment CBCT projections to represent the dynamic motion. In addition, it also serves to map future, intra-treatment x-ray projections into motion coefficients to solve instantaneous motion for real-time CBCT estimation.

*2.1.1 Learning task 1: dynamic CBCT reconstruction*

Since the DREME framework was built upon our previous work of dynamic CBCT reconstruction (Shao et al. 2024), we summarized the previous work below and highlighted the key differences in DREME, especially the components for the real-time motion estimation.

Our previous model (called PMF-STINR) takes spatiotemporal coordinates (i.e., $\boldsymbol{x}$ and *t*) as network inputs and generates the reference CBCT $\boldsymbol{I}_{ref}(\boldsymbol{x})$, MBCs $\boldsymbol{e}_i(\boldsymbol{x})$, and MBC scores $\boldsymbol{w}_i(t)$ at the query spatiotemporal coordinates. The reference CBCT $\boldsymbol{I}_{ref}(\boldsymbol{x})$ and MBC scores $\boldsymbol{w}_i(t)$ are represented by implicit neural representations (INRs) (Mildenhall et al. 2022, Tewari et al. 2022, Molaei et al. 2023), while the MBCs are similarly parametrized by B-spline interpolants as DREME. The INR learns a continuous mapping from the input spatial/temporal spaces to the corresponding image/score domains, using neural networks (i.e., multilayer perceptrons) as non-parametric universal function approximators. INR has demonstrated high learning efficiency (Mildenhall et al. 2022) and superior performance for representing complex functions, and has been applied to various medical image reconstruction and registration problems (Reed et al. 2021, Shen et al. 2022, Zha et al. 2022, Lin et al. 2023, Zhang et al. 2023). Before feeding the spatiotemporal coordinates into the spatial and temporal INRs, learnable hashing encoders (Muller et al. 2022) are employed to enable PMF-STINR to capture high-frequency details of the underlying functions. The training process of PMF-STINR was driven by a similarity loss function that matches the DRRs of the dynamic CBCTs with the cone-beam acquisition $p$, similar to the learning task 1 of DREME. However, since the time index $t$ of PMF-STINR only enumerates the acquisition order of the x-ray projections $p$, it lacks physical meaning and cannot be extended to infer motion in projections acquired outside of $p$. To directly infer motion from all potential cone-beam projections, in DREME we replaced the temporal INR and its hash encoder by a CNN-based motion encoder that takes an x-ray projection ($p$) as input and outputs the MBC scores $w_i(p)$. Correspondingly, the CNN motion encoder acts upon true physical signals (an x-ray projection $p$), and can both solve motion in the pre-treatment CBCT projections and infer motion in intra-treatment, real-time x-ray projections. As illustrated in Fig. 1, the CNN-based motion encoder contains six layers of 2D convolution layers with 3×3 kernels. The feature maps of the layers comprise 2, 4, 8, 16, 32, 32 channels, respectively. Each of the convolution layers is followed by a batch normalization layer and a rectified linear unit function (ReLU). After the last ReLU, the feature maps are flattened and processed by a linear layer with nine outputs. Each output channel represents an MBC score $w_{i,k}(p)$, where $i$=1, 2, 3 and $k$=$x$, $y$, $z$. The subscripts $i$ and $k$ represent the index of the MBC levels (1, 2, and 3) and the index of the Cartesian components ($x$, $y$, and $z$), respectively

Two similarity loss functions were used to drive the training process of learning task 1 of DREME. They were defined on the image- and projection-domain, respectively. The image-domain loss served as the training label to warm start the spatial INR of DREME, and the projection-domain loss drove the joint training process of the reference CBCT and the data-driven motion model. Specifically, the image-domain loss function $L_{sim}^{im}$ was defined as the mean squared error between the spatial INR $I_{ref}(x)$ and an approximate CBCT $I_{app}(x)$:

$$L_{sim}^{im} = \frac{1}{N_{voxel}} \sum_{l=1}^{N_{voxel}} |I_{ref}(x_l) - I_{app}(x_l)|^2, \tag{3}$$

Where $l$ is the index for voxels and $N_{voxel}$ is the number of voxels in the reference CBCT. On the other hand, the projection-domain loss function $L_{sim}^{prj}$ compared the DRRs with the corresponding cone-beam projections:

$$L_{sim}^{prj} = \frac{1}{N_{batch} N_{pixel}} \sum_{t \in batch} \sum_{N_{pixel}} |\mathcal{P}[I(x,p)] - p|^2, \tag{4}$$

where $N_{batch}$ and $N_{pixel}$ respectively are the sample number in a batch and the pixel number of a projection, and $\mathcal{P}$ denotes the cone-beam projector that simulates DRRs from the dynamic CBCT $I(x,p)$. We set $N_{batch} = 32$ to balance the training speed and accuracy.

In addition to the similarity losses, a number of regularization losses were implemented to regularize the spatiotemporal reconstruction problem. The first regularization loss was the total variation (TV) loss on the spatial INR-reconstructed reference CBCT to suppress high-frequency image noise while preserving anatomy edges:

$$L_{TV} = \frac{1}{N_{voxel}}\sum_l |\nabla I_{ref}(x_l)|, \qquad (5)$$

where $\nabla$ denotes the gradient operator. In addition, regularization losses were implemented to regularize the data-driven motion model. The second regularization loss promoted the ortho-normality of the MBCs:

$$L_{MBC} = \frac{1}{9}\sum_{k=x,y,z}\sum_{i=1}^{3}\left(\left|\|e_{i,k}\|^2 - 1\right|^2 + \sum_{j=i+1}^{3}|e_{i,k}\cdot e_{j,k}|^2\right). \qquad (6)$$

The normalization constraint (i.e., the first term in the parenthesis in Eq. (6)) removes the ambiguity of the spatiotemporal decomposition Eq. (2) of the low-rank motion model, and the orthogonal constraint (i.e., the second term in the parenthesis in Eq. (6)) removes the ambiguity of the MBCs between different levels. The third regularization loss was the zero-mean score loss on the MBC scores $w_i(p)$:

$$L_{ZMS} = \frac{1}{9}\sum_{k=x,y,z}\sum_{i=1}^{3}\left|\frac{1}{N_p}\sum_p w_{i,k}(p)\right|^2, \qquad (7)$$

where $N_p$ denotes the number of projections. The zero-mean score loss was enforced to remove the constant baseline of $w_i(p)$.

Finally, a DVF self-consistency regularization loss was implemented. The self-consistency regularization enforces that the deformed reference CBCTs $I(x, p)$ followed by the inverse DVFs $d^{-1}(x, p)$ should return to the reference CBCT, which further conditions the ill-posed reconstruction problem:

$$L_{SC} = \frac{1}{N_{batch}}\sum_{t\in batch}\frac{1}{N_{voxel}}\sum_l |I'(x_l, p) - I_{ref}(x_l)|^2, \qquad (8)$$

where $I'(x, p) = I(x + d^{-1}(x, p), p)$. The inverse DVFs were calculated using the iterative algorithm (Chen et al. 2008) with three iterations.

In summary, the first learning task of DREME is similar to PMF-STINR, but with the temporal INR replaced by a CNN-based motion encoder. In addition, a zero-mean score loss on the motion coefficients and a DVF self-consistency regularization loss were added to further condition the learning. For more details of PMF-STINR, please refer to our previous publication (Shao et al. 2024).

*2.1.2 Learning task 2: real-time motion estimation and CBCT imaging*

Although the CNN-based motion encoder is able to infer motion directly from cone-beam projections and can readily represent both pre-treatment dynamic motion and intra-treatment real-time motion, the motion encoder trained using only the pre-treatment CBCT projections may simply learn to memorize the correspondence between the training projections and the corresponding scores as in the first learning task, and will not generalize to new, unseen motion observed in arbitrarily-angled intra-treatment projections. To train a motion- and angle-robust CNN, we need a second learning task to further enhance the CNN's capability in real-time motion and CBCT estimation. The second learning task further trains the CNN to estimate motion in real time from a single x-ray projection at an arbitrary projection angle, where the x-ray projection is augmented with motion different from that observed in the pre-treatment CBCT scan.

To simulate random deformable motion to further train the CNN, deformable augmentation was used to generate motion states unseen in the pre-treatment acquisition. The augmentation was implemented by randomly resampling the solved MBC scores $w_{i,k}(p)$ from the dynamic CBCT reconstruction for each projection, using the following equation:

$$w_{i,k}^{'}(p) = r_1(p) \times r_2(i,k) \times w_{i,k}(p), \quad (9)$$

where $r_1 \in [0.6, 2.0]$ and $r_2 \in [0.8, 1.2]$ are uniformly distributed random variables. Two random variables were used to instill independent randomness between different levels $i$ and Cartesian directions $k$ (through the level/direction-specific factors $r_2(i,k)$), while still maintaining a certain degree of motion correlation among the MBC scores of each instance (through the same overall scaling factor $r_1(p)$ for each projection). Based on the augmented MBC scores, the DVFs are randomized (Eq. 2), resulting in various deformation-augmented CBCTs (Eq. 1). From the deformation-augmented CBCTs, we generated on-the-fly DRRs using a cone-beam projector during the training, and input them into the motion encoder CNN to infer the MBC scores $w_{k,i}^{''}(p)$. The augmentation loss was then defined as

$$L_{Aug} = \frac{1}{9N_{batch}} \sum_{t \in batch} \sum_{k=x,y,z} \sum_{i=1}^{3} \left| w_{i,k}^{''}(p) - w_{i,k}^{'}(p) \right|^2. \quad (10)$$

During the network training, we avoided using intermediate motion models still in training for deformation augmentation (Table 1). Instead, for the motion models still in training, we did not perform additional deformation augmentation (as in Eq. 9). Instead, we only re-generated DRRs at random angles (angle augmentation only) from the solved dynamic CBCTs, and fed them into the CNN motion encoder to infer the MBC scores to compute the loss in Eq. 10.

2.2 A progressive training strategy

For DREME training, we followed our previous works (Zhang et al. 2023, Shao et al. 2024) by designing a progressive multiresolution training strategy to improve learning efficiency and avoid overfitting. The strategy progressively increases the learning complexity under two spatial resolutions (Table 1). The total loss function $L_{tot}$ was a weighted sum of the aforementioned loss functions, with corresponding weighting factors specified in Table 1.

**Table 1**. The progressive multiresolution training strategy with different loss functions. The symbol ✓ indicates that the corresponding loss function was used in the specific training stage.

| Training stage | | I-a | I-b | I-c | II-a | II-b | II-c | II-d |
|---|---|---|---|---|---|---|---|---|
| Number of epochs | | 400 | 700 | 1700 | 1000 | 1000 | 1000 | 1000 |
| Spatial resolution (mm³) | | | 4×4×4 | | | | 2×2×2 | |
| Spatial INR learning rate | | 4×10⁻⁴ | 4×10⁻⁵ | 1×10⁻⁵ | 1×10⁻³ | 4×10⁻⁴ | 1×10⁻⁴ | 0. |
| Motion encoder learning rate | | 0. | 0. | 1×10⁻³ | 0. | 0. | 1×10⁻⁴ | 1×10⁻⁴ |
| B-spline interpolant learning rate | | 0. | 0. | 1×10⁻³ | 0. | 0. | 1×10⁻⁴ | 0. |
| Loss function | Weighting factor λ | | | | | | | |
| Image-domain similarity loss Eq. (3) | 1. | ✓ | | | ✓ | | | |
| Projection-domain similarity loss | 1. | | ✓ | ✓ | | ✓ | ✓ | |

| | | | | | | |
|---|---|---|---|---|---|---|
| Eq. (4) TV regularization | $2\times10^{-4}$ | ✓ | ✓ | ✓ | ✓ | |
| Eq. (5) MBC regularization | 1. | | ✓ | | ✓ | |
| Eq. (6) Zero-mean score regularization | $1\times10^{-3}$ | | ✓ | | ✓ | |
| Eq. (7) DVF self-consistency regularization | $1\times10^{3}$ | | ✓ | | ✓ | |
| Eq. (8) Motion/angle augmentation loss Eq. (10) | $1\times10^{-4}$ | | | ✓ (angle augmentation only) | ✓ (angle augmentation only) | ✓ (motion + angle augmentation) |

Remarks on the training strategy are in order: **(a)** The low-resolution reconstruction contained the training stage I with an isotropic 4×4×4 mm³ resolution (for intermediate CBCT inference from the spatial INR), and the resolution was doubled at the stage II (2×2×2 mm³). Due to the GPU memory limits, we were unable to proceed to a higher resolution, though this constraint is not a fundamental limitation of the proposed framework. **(b)** Both low- and high-resolution stages followed a parallel training structure that each started from image-domain similarity loss (Eq. (3)) and then switched to projection-domain similarity loss (Eq. (4)). The motion encoder and B-spline interpolants entered the training at later sub-stages (I-c and II-c). **(c)** The approximate CBCT $I_{app}(x)$ at low-resolution training was obtained from the Feldkamp-Davis-Kress (FDK) reconstruction (Feldkamp et al. 1984) of all the x-ray projections within $p$. At the stage of the high-resolution reconstruction, it was obtained by the tri-linear up-sampling from the solved $I_{ref}(x)$ at the low-resolution training stage. **(d)** Zero learning rate meant the learnable parameters in the corresponding network component were frozen. **(e)** The second learning task (Fig. 1) was introduced at stage II (from II-b). When the spatial INR and the motion model were still in training (non-zero learning rates), we did not perform deformation augmentation, and only re-generated DRRs from the pre-treatment x-ray projections with new, random angles (angle augmentation only) to train the CNN motion encoder. The last stage (stage II-d) was dedicated to the second learning task with both motion and angle augmentation enabled, and the spatial INR and the B-spline motion model frozen (learning rates set to 0). **(f)** The values of the weighting factors were determined by empirical searching.

2.3 Evaluation datasets and schemes

We evaluated DREME using the extended cardiac torso (XCAT) digital phantom (Segars et al. 2010) and a multi-institutional lung patient dataset. The XCAT simulation study provided 'ground truth' for quantitative evaluation, thus allowing us to assess the model and training strategy designs and optimize network hyper-parameters. The patient study allowed us to evaluate the clinical application potential of DREME. Given the different natures of the two datasets, we discuss them separately in the following subsections.

*2.3.1 XCAT simulation study*

We simulated an XCAT phantom to cover the thorax and upper abdomen region, with a volume of 200×200×100 voxels and a 2×2×2 mm$^3$ voxel size. A spherical lung tumor with a 30-mm diameter was inserted in the lower lobe of the right lung as a motion tracking target. Seven respiratory motion trajectories (X1-X7) were simulated (Fig. 2), featuring motion scenarios combining variations in breathing amplitudes, frequencies, and/or patterns. Different motion trajectories were used to evaluate DREME's generalizability to different motion scenarios, especially when there is a discrepancy between the training and testing scenarios. After generating XCAT volumes for these scenarios, we simulated x-ray projections $p$ from them as pre-treatment cone-beam scans. The projections were generated in full-fan mode, including 660 projections covering a 360° gantry angle. The scan time was 60 seconds (i.e., 11 frames per second) to mimic the clinical setting. Each projection spanned 256×192 pixels with a 1.55×1.55 mm$^2$ pixel resolution. In addition to the pre-treatment scans, we also simulated projections on each motion scenario to serve as intra-treatment scans to test the real-time imaging capability of DREME. To test the angle-agnostic capability of the trained CNN motion encoder, the second sets of cone-beam projections were simulated with the gantry angle rotated by 90.27° respect to the pre-treatment sets, to ensure that no two x-ray projections in the training and testing sets share the same motion state and the projection angle. We used the tomographic package ASTRA toolbox to simulate the cone-beam projections (van Aarle et al. 2016). The same toolbox was also used in the DREME network as the cone-beam projector.

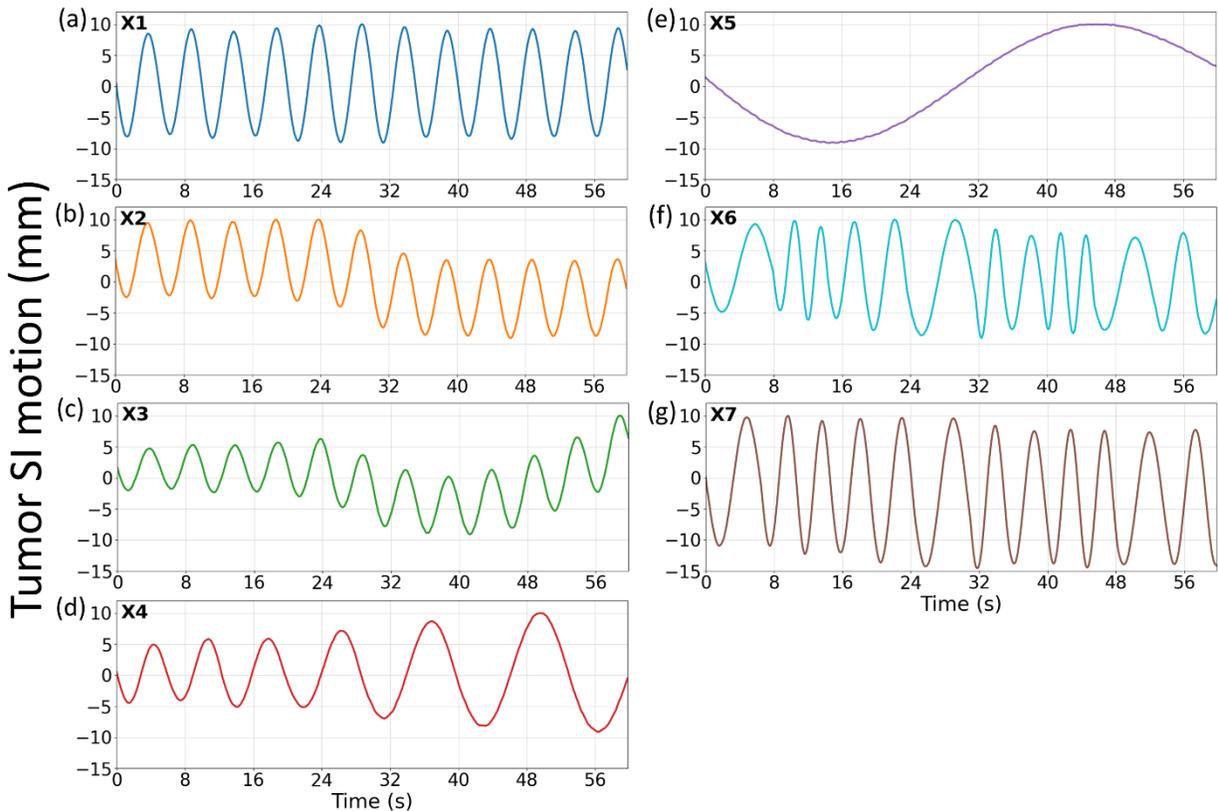

**Figure 2**. Lung tumor motion trajectories along the superior-inferior (SI) direction in the XCAT simulation study. Trajectories X1-X6 shared the same maximum motion ranges in the SI direction, whereas the SI range of X7 were extended to evaluate the robustness of DREME models trained using the other scenarios (X1-X6).

As a 'one-shot' learning technique, DREME was separately trained on each of the first six motion scenarios (X1-X6) and tested on all scenarios for real-time motion estimation (X1-X7), yielding six groups of results. This cross-scenario testing evaluates the robustness of DREME to testing scenarios very different from the training (Fig. 2). The performance of DREME was evaluated based on the image quality of the solved real-time CBCTs $I(x,p)$ and the accuracy of lung tumor localization. The image quality was evaluated via the mean relative error metric (RE):

$$\text{RE} = \frac{1}{N_p} \sum_p \sqrt{\frac{\sum_{l=1}^{N_{voxel}} \|I(x_l,p) - I^{gt}(x_l,p)\|^2}{\sum_{l=1}^{N_{voxel}} \|I^{gt}(x_l,p)\|^2}}, \quad (11)$$

where $I^{gt}(x,p)$ is the 'ground-truth' CBCT. The structural similarity index measure (SSIM) was also evaluated between the reconstructed and 'ground-truth' real-time CBCTs. The tumor localization accuracy was evaluated by contour-based metrics, including the tumor center-of-mass error (COME) and Dice similarity coefficient (DSC). We contoured the lung tumor from the reconstructed reference CBCTs, and then propagated the reference tumor contours to other instances using solved real-time DVFs and compared with the 'ground-truth' contours.

*2.3.2 Patient study*

The patient dataset included eight lung patients from three institutes. Three patients (P1-P3) were from the MD Anderson Cancer Center (Lu et al. 2007), one patient (P6) from the UT Southwestern Medical Center (Shao et al. 2024), and four patients (P4-P5, P7-P8) from the SPARE challenge dataset (Shieh et al. 2019). These patients were selected with observed respiration-induced motion in the thoracic-abdominal region. The scans were under various acquisition protocols including full-/half-fan scans, different kVp/mA/mS, and varying cone-beam geometries (i.e., source-to-axis distance/source-to-detector distance) etc. The acquisition protocols, sizes of the CBCT volumes, and other details of the dataset can be found in our previous work (Table 1 in (Shao et al. 2024)). As all cases had only a single pre-treatment scan, the cone-beam projections were divided into a training set and a testing set. For P1-P3 and P6, every third projection from their pre-treatment scan was selected for model training, and the remaining two-thirds of the projections were used for model testing. For the SPARE challenge patients (P4-P5, P7-P8), the dataset already included down-sampled scans for the spare-view reconstruction challenge (Shieh et al. 2019). Thus, we trained DREME on the down-sampled scans and tested it on the fully-sampled sets, excluding the projections already in the training sets.

As 'ground-truth' data were unavailable for the patient study, to evaluate the accuracy of motion estimation, we re-projected the reconstructed CBCT volumes into 2D DRRs and compared them with the acquired projections in the testing set, via motion features tracked by the Amsterdam Shroud (AS) method (Zijp et al. 2004). Details of the AS method were provided in our previous work (Shao et al. 2024). In brief, the intensity gradients along the vertical axis (i.e., the SI direction) were calculated for both cone-beam projections and DRRs to highlight anatomic landmarks with high contrast edges (e.g., diaphragm). These intensity gradients were integrated along the horizontal axis over a region of interest that exhibited clear motion-induced intensity variations to form an AS image. Finally, the AS images were post-processed to enhance their contrast for motion trace extraction, and the localization accuracy was calculated by the difference between the extracted traces from the cone-beam projections and from the DRRs. In addition, Pearson correlation coefficients between the extracted traces were calculated.

2.4 Comparison and ablative studies

We compared DREME with a principal component analysis (PCA)-based 2D3D deformable registration technique (Li et al. 2011, Shao et al. 2022) on the XCAT dataset. The 2D3D registration solves a 3D DVF by matching DRRs of the deformed/registered anatomy with acquired x-ray projections. This technique incorporates patient-specific prior knowledge to alleviate the instability of the inverse problem. We used PCA-based motion models for the 2D3D registration. Two variants of the motion models were developed, depending on the sources from which the motion models were derived. The PCA motion model can be derived from a 4D-CT scan during treatment planning, which can potentially introduce biases (Sec. 1). Alternatively, the motion model can be directly derived from the pre-treatment cone-beam scan (4D-CBCT). However, 4D reconstruction requires motion sorting/binning of the projections according to the respiratory phases, typically resulting in an under-sampled reconstruction. We sorted the pre-treatment scan projections into 10 respiratory phases and obtained intra-phase DVFs by registering the 4D-CBCT to the end-of-exhale phase CBCT, and then performed PCA on the intra-phase DVFs to derive principal motion components for the 2D3D deformable registration. DREME was compared with both variants, which were called 2D3D$_{PCA-4DCT}$ and 2D3D$_{PCA-4DCBCT}$, respectively, according to the source of the PCA motion model. Wilcoxon signed-rank tests were used to evaluate the significance level of observed differences.

As discussed in Sec. 2.1.1 and shown in Table 1, a number of loss functions were used to regularize the inverse problem and achieve a stable motion model. To investigate the benefits and roles of these regularization terms, an ablation study was conducted on the XCAT dataset by removing different loss functions in the training process. We removed the zero-mean score loss $L_{ZMS}$, DVF self-consistency loss $L_{SC}$, motion/angle augmentation loss ($L_{Aug}$), and evaluated the impacts on the image quality and motion tracking accuracy. These variants of the DREME framework in the ablation study are called DREME$_{WO-ZMS}$, DREME$_{WO-SC}$, and DREME$_{WO-Aug}$, where the subscript indicates which loss functions were removed. Wilcoxon signed-rank tests were used to evaluate the significance level of observed differences.

## 3. Results

3.1 The XCAT study results

Figure 3 compares the reference CBCTs used in the 2D3D-based methods and those reconstructed by DREME for the six motion scenarios (X1-X6) in the axial and coronal views. The reference CBCTs of 2D3D$_{PCA-4DCT}$ and 2D3D$_{PCA-4DCBCT}$ correspond to the end-of-exhale phase. 2D3D$_{PCA-4DCT}$ used the end-of-exhale phase image from a prior 4D-CT scan, and the 4D-CT-derived PCA motion model was used in 2D3D deformable registrations for all motion scenarios (X1-X6). For 2D3D$_{PCA-4DCBCT}$, the reference CBCTs were reconstructed from the phase-sorted pre-treatment scans, which were severely affected by under-sampling artifacts (~66 projections for the end-of-exhale phase). The X5 scenario of slow breathing (or, equivalently, fast gantry rotation) also suffered from the issue of limited-angle reconstruction, as the end-of-exhale projections were confined to a narrow range of projection angles. These artifacts posed substantial challenges for the 2D3D-based registration technique.

## (a) 2D3D$_{PCA-4DCT}$

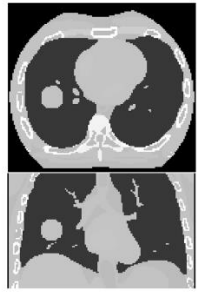

X1    X2    X3    X4    X5    X6

## (b) 2D3D$_{PCA-4DCBCT}$

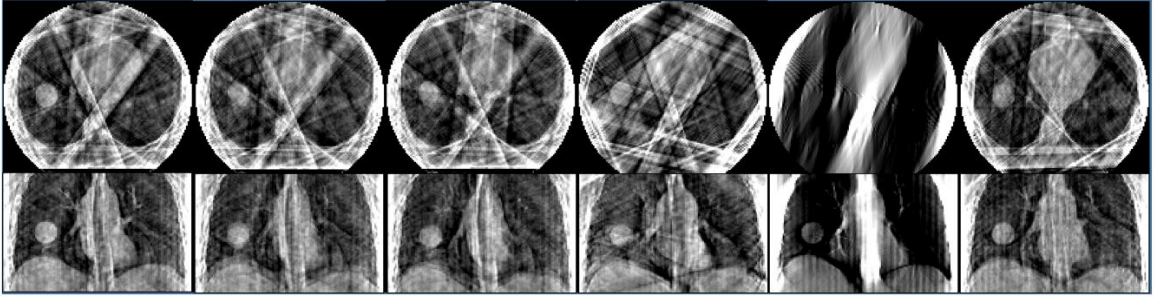

## (c) DREME

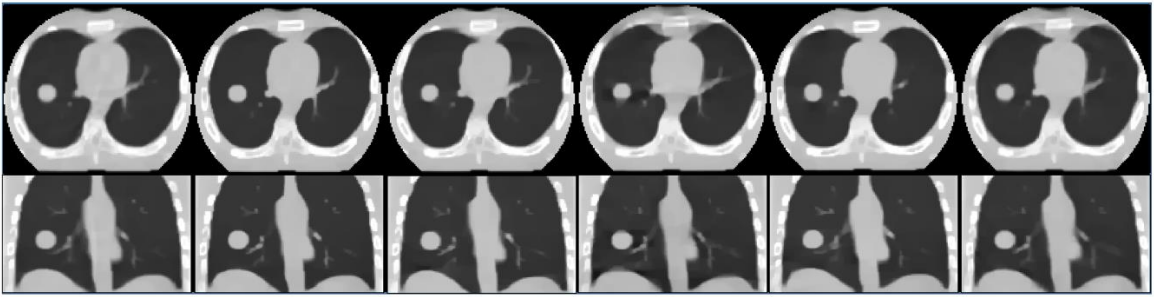

**Figure 3**. Comparison of reconstructed reference CBCTs in the comparison study. 2D3D$_{PCA-4DCT}$ assumed the existence of an artifact-free pre-treatment 4D-CT scan and used the end-of-exhale phase as the reference anatomy. 2D3D$_{PCA-4DCBCT}$ and DREME reconstructed the reference CBCTs using the pre-treatment scans, thus the anatomy is up-to-date and specific to daily motion scenarios. The reference CBCTs of 2D3D$_{PCA-4DCT}$ were reconstructed from the end-of-exhale phase after projection phase sorting, suffering from significant under-sampling and motion-related artifacts. DREME reconstructed the reference CBCTs with simultaneous motion estimation/compensation, resulting in minimal motion-related artifacts.

Tables 2 and 3 respectively summarize the mean relative error and mean SSIM from the comparison and ablation studies. All Wilcoxon signed-rank tests between DREME and other methods yielded p values $< 10^{-3}$. DREME outperformed the 2D3D-based techniques in terms of both relative error and SSIM. On the other hand, for the ablation study, DREME offered the best results among all, demonstrating the benefits of including various loss functions to condition the real-time imaging problem, especially the motion/angle augmentation losses used in the learning task 2.

**Table 2**. Mean relative error of the XCAT simulation study. The results for each training scenario (X1-X6) are presented as the mean and standard deviation (Mean±SD), averaged over all testing scenarios (X1-X7).

| Training scenario | Comparison study | | Ablation study | | | |
|---|---|---|---|---|---|---|
| | 2D3D$_{PCA-4DCT}$ | 2D3D$_{PCA-4DCBCT}$ | DREME$_{WO-ZMS}$ | DREME$_{WO-SC}$ | DREME$_{WO-Aug}$ | DREME |
| X1 | | 0.575±0.053 | 0.164±0.014 | **0.162±0.016** | 0.198±0.047 | **0.162±0.015** |
| X2 | | 0.591±0.066 | 0.166±0.016 | 0.164±0.016 | 0.214±0.056 | **0.158±0.018** |
| X3 | 0.192±0.050 | 0.617±0.067 | 0.169±0.022 | 0.168±0.020 | 0.223±0.060 | **0.165±0.022** |
| X4 | | 0.732±0.074 | 0.191±0.015 | 0.193±0.017 | 0.243±0.052 | **0.185±0.015** |
| X5 | | 1.328±0.379 | 0.173±0.021 | 0.177±0.021 | 0.261±0.049 | **0.172±0.019** |
| X6 | | 0.562±0.070 | **0.169±0.016** | 0.170±0.014 | 0.215±0.053 | 0.170±0.015 |

**Table 3**. Mean SSIM of the XCAT simulation study. The results for each training scenario (X1-X6) are presented as the mean and standard deviation (Mean±SD), averaged over all testing scenarios (X1-X7).

| Training scenario | Comparison study | | Ablation study | | | |
|---|---|---|---|---|---|---|
| | 2D3D$_{PCA-4DCT}$ | 2D3D$_{PCA-4DCBCT}$ | DREME$_{WO-ZMS}$ | DREME$_{WO-SC}$ | DREME$_{WO-Aug}$ | DREME |
| X1 | | 0.8500±0.0130 | 0.9810±0.0031 | 0.9813±0.0035 | 0.9728±0.0122 | **0.9815±0.0033** |
| X2 | | 0.8477±0.0161 | 0.9804±0.0035 | 0.9808±0.0035 | 0.9684±0.0145 | **0.9819±0.0039** |
| X3 | 0.9777±0.0109 | 0.8433±0.0131 | 0.9795±0.0051 | 0.9795±0.0048 | 0.9663±0.0161 | **0.9803±0.0053** |
| X4 | | 0.8029±0.0127 | 0.9743±0.0039 | 0.9737±0.0047 | 0.9596±0.0154 | **0.9759±0.0038** |
| X5 | | 0.6027±0.0165 | **0.9784±0.0046** | 0.9770±0.0046 | 0.9556±0.0138 | **0.9784±0.0042** |
| X6 | | 0.8549±0.0164 | **0.9794±0.0036** | 0.9793±0.0032 | 0.9678±0.0142 | 0.9791±0.0033 |

Tables 4 and 5 respectively summarize the mean tumor COME and DSC results. All Wilcoxon signed-rank tests of COME between DREME and the other methods yielded p values < 10$^{-3}$, except for DREME$_{WO-ZMS}$ (p = 0.007), and all Wilcoxon signed-rank tests of DSC between DREME and the other methods yielded p values < 10$^{-3}$. Even with a high-quality reference CBCT, it remained challenging for 2D3D$_{PCA-4DCT}$ to locate the tumors accurately. We found the error generally increased when the respiratory phase of real-time CBCTs moved away from the reference CBCTs (i.e., end-of-exhale phase), which indicates that the 2D3D-based registration algorithm may have been trapped at local minima when estimating motion with large amplitudes. 2D3D$_{PCA-4DCBCT}$ had very poor localization performance, as the reconstruction artifacts in the reference CBCTs propagated to the tumor localization errors. Overall, DREME achieved sub-voxel COME.

**Table 4**. Lung tumor center-of-mass error (COME, in mm) of the XCAT simulation study. The results of each training scenario show the mean and standard deviation (Mean±SD) averaged over all testing scenarios (X1-X7).

| Training scenario | Comparison study | | Ablation study | | | |
|---|---|---|---|---|---|---|
| | 2D3D$_{PCA-4DCT}$ | 2D3D$_{PCA-4DCBCT}$ | DREME$_{WO-ZMS}$ | DREME$_{WO-SC}$ | DREME$_{WO-Aug}$ | DREME |
| X1 | | 7.0±4.4 | 1.0±0.7 | 1.0±0.9 | 3.6±3.4 | **0.9±0.8** |
| X2 | | 7.6±4.6 | 1.1±0.8 | 1.6±0.8 | 5.0±3.8 | **1.0±0.7** |
| X3 | 3.1±3.5 | 8.4±4.7 | **1.2±0.9** | 1.3±1.0 | 5.5±4.2 | **1.2±1.0** |
| X4 | | 10.0±5.0 | **1.6±1.1** | 1.9±0.9 | 6.3±4.1 | 1.7±0.9 |
| X5 | | 15.4±6.7 | **1.3±1.1** | 1.5±1.2 | 7.5±4.0 | **1.3±1.0** |
| X6 | | 9.6±4.7 | **1.2±0.9** | 1.2±0.9 | 4.9±3.9 | 1.3±0.8 |

**Table 5**. Lung tumor Dice similarity coefficient (DSC) of the XCAT simulation study. The results of each training scenario show the mean and standard deviation (Mean±SD) averaged over all testing scenarios (X1-X7).

| Training scenario | Comparison study | | Ablation study | | | |
|---|---|---|---|---|---|---|
| | 2D3D$_{PCA-4DCT}$ | 2D3D$_{PCA-4DCBCT}$ | DREME$_{WO-ZMS}$ | DREME$_{WO-SC}$ | DREME$_{WO-Aug}$ | DREME |
| X1 | | 0.640±0.144 | 0.918±0.028 | **0.922±0.034** | 0.812±0.147 | 0.919±0.029 |
| X2 | | 0.590±0.124 | 0.902±0.029 | 0.920±0.029 | 0.757±0.166 | **0.922±0.028** |
| X3 | 0.850±0.135 | 0.596±0.159 | 0.900±0.032 | **0.913±0.037** | 0.731±0.181 | 0.910±0.035 |
| X4 | | 0.570±0.133 | 0.861±0.035 | 0.881±0.033 | 0.692±0.178 | **0.884±0.032** |
| X5 | | 0.334±0.184 | 0.895±0.042 | 0.885±0.045 | 0.641±0.175 | **0.888±0.037** |
| X6 | | 0.478±0.178 | 0.892±0.029 | **0.910±0.032** | 0.754±0.169 | 0.902±0.031 |

Figure 4 presents the solved tumor motion trajectories in the comparison study. In this figure, DREME was trained on the X3 scenario and tested across all scenarios (X1-X7). Overall, DREME showed more accurate and stable performance. In contrast, 2D3D$_{PCA-4DCT}$ exhibited significant deviations in the SI direction, particularly near the end-of-inhale phase. Additionally, 2D3D$_{PCA-4DCBCT}$ showed most unstable tracking accuracy, due to its low-quality motion models. Figure 5 presents two scenarios (X5 and X6) of solved real-time CBCTs with tumor contours in the coronal and sagittal views and compares them with the corresponding 'ground-truth' CBCTs. The training scenario was X3. The real-time CBCTs solved by DREME matched well with the 'ground-truth'.

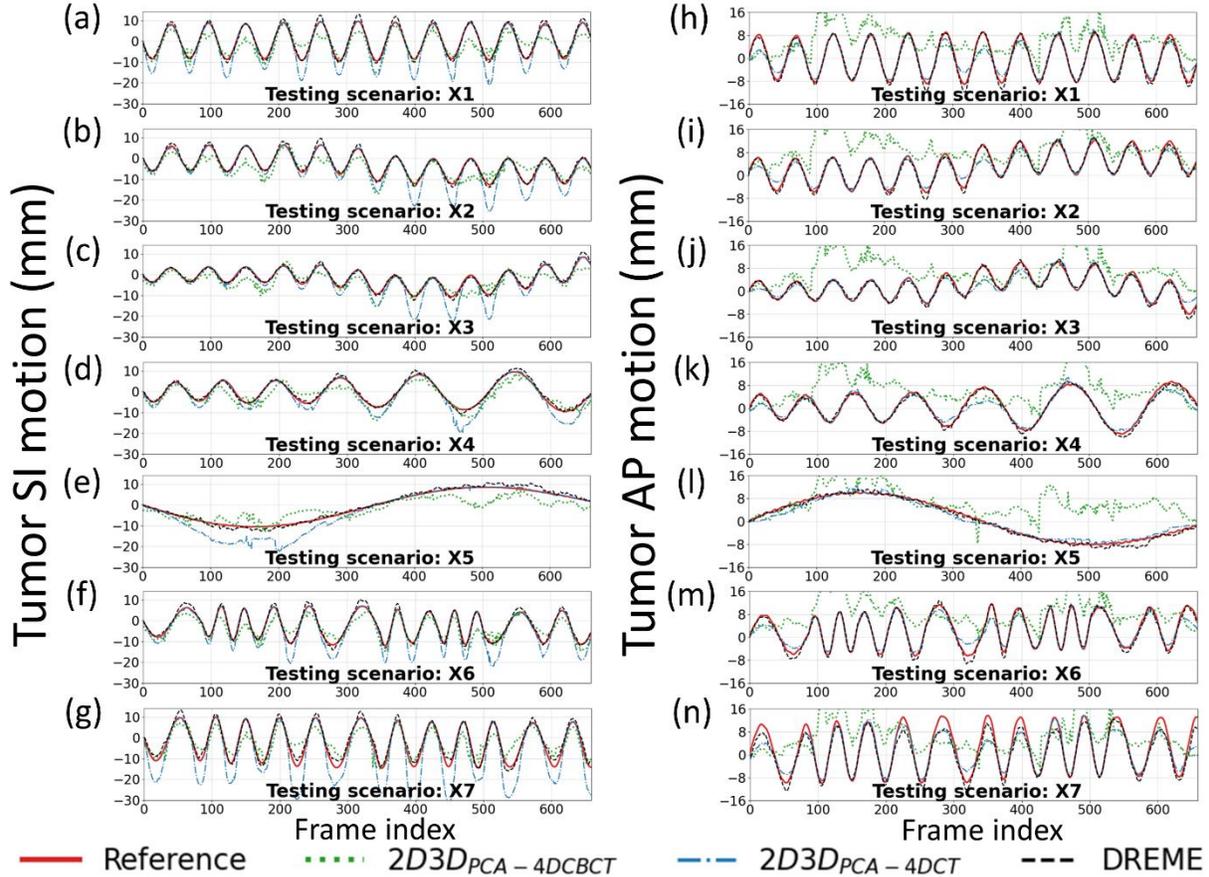

**Figure 4**. Tumor motion trajectories of the comparison study. The first and second columns respectively presents the comparison of the solved tumor motion for the X1-X7 scenarios in the SI and AP directions between 2D3D$_{PCA-4DCBCT}$, 2D3D$_{PCA-4DCT}$, DREME, and the 'ground-truth' reference. DREME was trained on the X3 scenario.

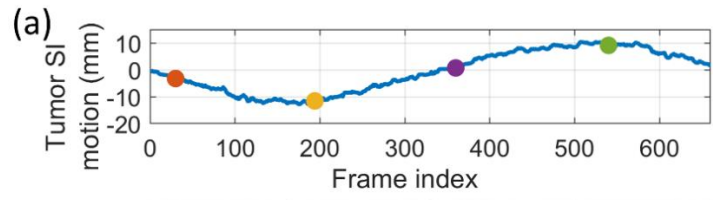
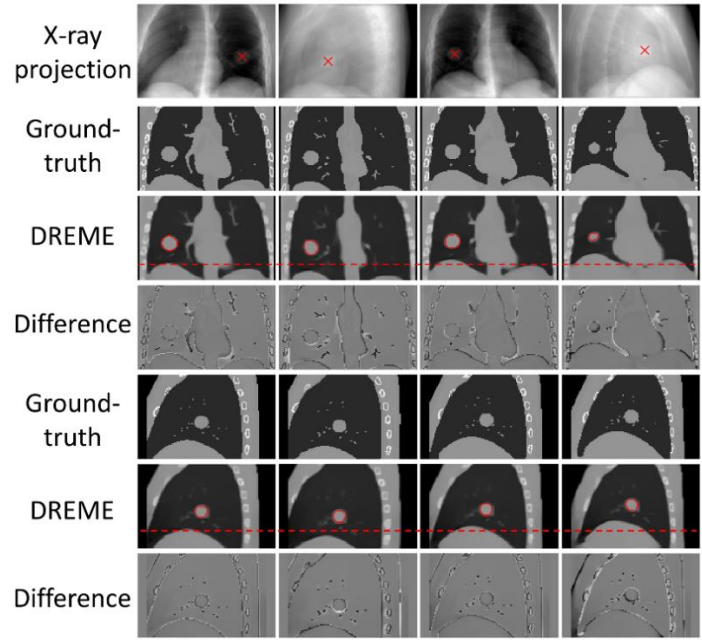
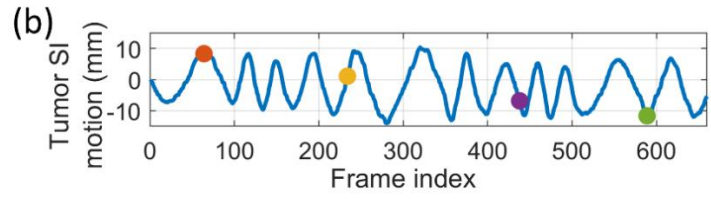
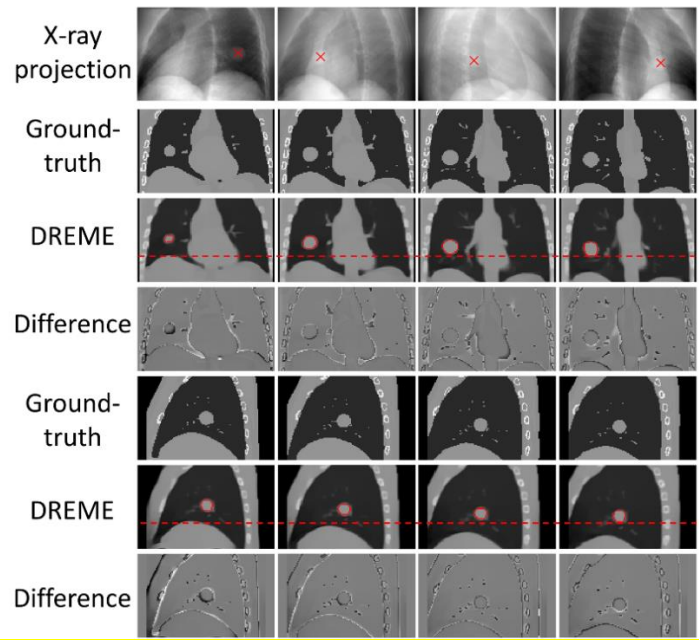

**Figure 5**. Examples of real-time CBCTs and lung tumor motion solved by DREME for the XCAT study: (a) X5 scenario and (b) X6 scenario. The training scenario was X3. The first row shows the tumor motion curves along the SI direction, with the dots indicating the time points selected for plotting. The second raw presents the onboard x-ray projections at the selected time points, and the "×" symbols indicate the solved tumor center-of-mass positions projected onto the projections. In the following rows, real-time CBCTs of the selected time points are compared against the 'ground-truth' CBCTs, with the difference images calculated. The estimated tumor contours (red) are also presented for each selected time points.

3.2 The patient study results

For the patient study, the localization error and Pearson correlation of all patients are summarized in Table 6. For all patients, the mean localization error is below 2.6 mm in the projection domain, and the estimated motion traces show high correlations (>0.92) with the reference traces. We note that the errors were evaluated in the projection domain, and the corresponding errors in the image domain would be scaled down by a factor of ~1.5 to achieve sub-voxel localization accuracy. Figure 6 presents two examples (P1 and P2) of solved real-time CBCTs in the coronal and sagittal views.

**Table 6**. Mean localization accuracy and Pearson correlation coefficient for the SI motion trajectories extracted from the AS images for the patient study. The results are presented in terms of the mean and standard deviation (Mean±SD) averaged over all testing projections for each patient.

| Patient ID | Pearson correlation coefficient (SI trajectory) | Localization error (mm) |
|---|---|---|
| P1 | 0.946 | 1.5±1.4 |
| P2 | 0.967 | 2.1±1.9 |
| P3 | 0.919 | 1.4±1.4 |
| P4 | 0.949 | 1.7±1.4 |
| P5 | 0.964 | 2.6±2.5 |
| P6 | 0.979 | 2.0±1.5 |
| P7 | 0.984 | 2.1±1.9 |
| P8 | 0.993 | 1.2±1.0 |

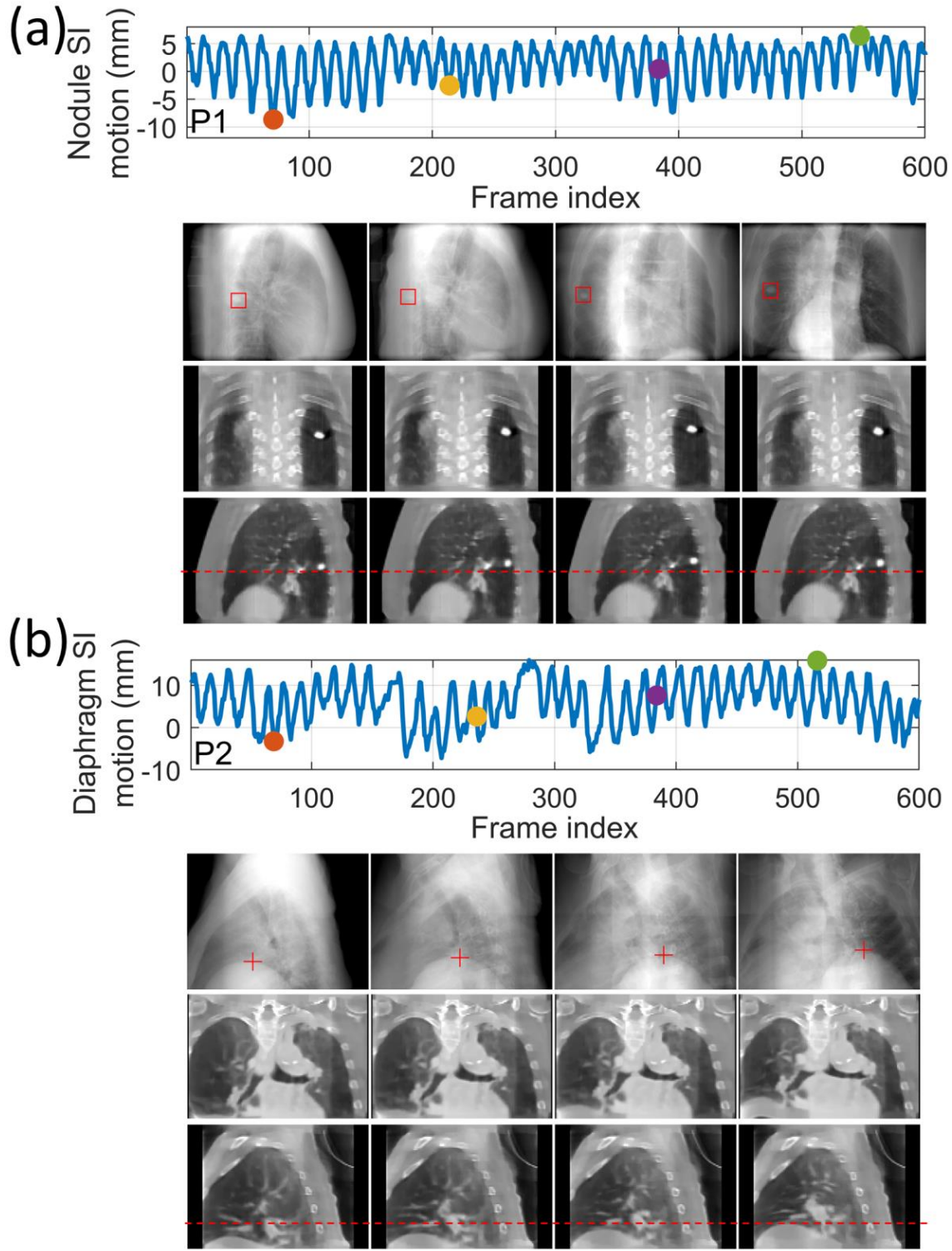

**Figure 6**. Real-time CBCTs and motion trajectories estimated by DREME for (a) P1 and (b) P2 of the patient study. The first row shows the SI motion trajectory of the tracking landmark (P1: lung nodule; P2: diaphragm apex), with the dots indicating the acquisition time points selected for plotting. The second raw presents the onboard x-ray projections at the selected time points, and the "□" and "+" symbols indicate the projected landmark positions in the projections. In the following rows, real-time CBCTs at the selected time points are presented in the coronal and sagittal views.

## 4. Discussion

In this work we proposed a framework called DREME for real-time CBCT imaging and motion estimation in radiotherapy. The framework adopts a dual-task learning strategy (Fig. 1) that incorporates a CNN-based real-time imaging method into a dynamic CBCT reconstruction workflow. The dynamic CBCT reconstruction solves a dynamic sequence of CBCTs from a standard pre-treatment scan to extract the latest patient anatomy and motion model, which is subsequently used for real-time CBCT motion estimation. The reconstruction algorithm is based on a joint deformable registration and reconstruction approach, and derives the motion model in a data-driven manner without relying on motion surrogate signals or patient-specific prior knowledge, eliminating the potential biases and uncertainties from these sources. To achieve a high-quality and realistic motion model, specially-designed regularization terms (Sec. 2.1.1) and a progressive multiresolution training strategy (Sec. 2, Table 1) were developed. With the lightweight CNN-based motion encoder of DREME, its inference time for estimating a 3D DVF was approximately 1.5 ms, fulfilling the temporal constraint of real-time imaging (Keall et al. 2021).

DREME was evaluated by an XCAT simulation study and a multi-institutional dataset of eight lung patients. The XCAT results showed that DREME can capture various types of motion scenarios differing from the training scenario (Fig. 3), and achieved a mean tumor COME of 1.2±0.9 mm (Table 4). For the patient study, DREME achieved a mean localization error of 1.8±1.6 mm with high Pearson correlation coefficients (Table 6). The comparison study clearly demonstrated that DREME is superior to the conventional 2D3D registration-based methods (Tables 2-5 and Fig. 4), and also illustrated the challenges of deriving high-quality motion models from standard pre-treatment scans by 4D-CBCTs. The ablation study showed that DREME's regularization terms enhanced the quality of the motion model (Tables 2-5).

Contrary to other DL-based approaches for real-time imaging, DREME does not require a large dataset or a high-quality prior motion model for training and data augmentation. Instead, DREME solves dynamic CBCTs and derives data-driven motion models in a "one-shot" fashion from pre-treatment cone-beam scans. Consequently, DREME does not suffer from generalizability and stability issues, and can be utilized under various CBCT acquisition protocols, as demonstrated in the patient studies (Fig. 6 and Table 6). Unlike 4D-CT/CBCT, which over-samples anatomic information to achieve artifact-free 4D reconstructions, DREME solves the motion model from standard 3D scans, avoiding additional radiation exposure to patients. The patient study demonstrated that DREME can extract an accurate motion model from sparsely-sampled projection sets, which can further reduce patient radiation dose.

As DREME relies on data-driven motion models for real-time motion estimation, the quality of the motion model directly influences the accuracy of the motion estimation. We observed a correlation between the quality of solved pre-treatment dynamic CBCTs and the performance of real-time motion estimation. Therefore, a direction for future work is to further improve the quality and robustness of the reconstructed CBCTs and the data-driven motion models. A major limitation of the DREME framework is the time for model training. The training time of DREME (based on CBCTs of 200×200×100 voxels) was approximately two hours on an Nvidia RTX 4090 GPU card. Several approaches could potentially be employed to accelerate the training process. Since the DREME network was trained from scratch for each projection set, transfer learning could reduce the training time. For example, the CNN-based motion encoder could be replaced by a pre-trained network (e.g., ResNet), and then fine-tuned for each case. Alternatively, one could pre-train the motion encoder using the cone-beam scan from a previous fraction/treatment, then fine tune the motion encoder using the current pre-treatment scan of the same patient, provided no dramatic variations in patient anatomy and motion. In addition, the cone-beam projector from the ASTRO toolbox parallelizes DRR simulations across multiple gantry angles, whereas DREME only requires a single DRR at each gantry angle. Consequently, DREME sequentially simulated

DRRs for the dynamic sequence to calculate the similarity loss. A more efficient parallelization scheme for the cone-beam projector tailored to the DREME framework could also speed up the model training. Furthermore, a more efficient approach for evaluating projection-domain similarity loss could also be employed. Currently, all DRR pixels were computed to quantify the similarity between the DRRs and x-ray projections. However, pixels do not play an equal role in similarity quantification. For instance, the upper thorax typically exhibits smaller amplitude and simpler pattern of respiratory motion than the lower thorax and the diaphragm region. Therefore, a better sampling scheme could be designed to increase the sampling rate of pixels exhibiting large and complicated motion. Since the above approaches require major modifications of the algorithms in the current workflow and are beyond the scope of the study, we leave these possibilities for future investigations.

## 5. Conclusion

We proposed a joint framework for real-time CBCT imaging and motion estimation. The proposed method reconstructed a dynamic sequence of CBCTs from a pre-treatment scan, and subsequently used the latest anatomy and motion model for real-time motion estimation to avoid potential biases associated with patient-specific prior knowledge. The results demonstrated that the framework can yield high-quality motion models and achieved sub-voxel localization accuracy in both simulation and real patient studies. The framework can have broad applications in radiotherapy, as the derived real-time motion allows potential real-time tumor tracking, dose accumulation, and adaptive radiotherapy.

**Acknowledgements**

The study was supported by funding from the National Institutes of Health (R01 CA240808, R01 CA258987, R01 CA280135, R01 EB034691), and from Varian Medical Systems. We would like to thank Dr. Paul Segars at Duke University for providing the XCAT phantom for our study.

**Ethical statement**

The MDACC dataset used in this study was retrospectively collected from an IRB-approved study at MD Anderson Cancer Center in 2007. The UTSW dataset was retrospectively collected from an approved study at UT Southwestern Medical Center on August 31, 2023, under an umbrella IRB protocol 082013-008 (Improving radiation treatment quality and safety by retrospective data analysis). This is a retrospective analysis study and not a clinical trial. No clinical trial ID number is available. In both datasets, individual patient consent was signed for the anonymized use of the imaging and treatment planning data for retrospective analysis. These studies were conducted in accordance with the principles embodied in the Declaration of Helsinki.

## References

Bernier, J., E. J. Hall and A. Giaccia (2004). "Radiation oncology: a century of achievements." Nat Rev Cancer **4**(9): 737-747.

Chen, M., W. Lu, Q. Chen, K. J. Ruchala and G. H. Olivera (2008). "A simple fixed-point approach to invert a deformation field." Med Phys 35(1): 81-88.

Cui, Y., J. G. Dy, G. C. Sharp, B. Alexander and S. B. Jiang (2007). "Multiple template-based fluoroscopic tracking of lung tumor mass without implanted fiducial markers." Physics in Medicine and Biology 52(20): 6229-6242.

Feldkamp, L. A., L. C. Davis and J. W. Kress (1984). "Practical Cone-Beam Algorithm." Journal of the Optical Society of America a-Optics Image Science and Vision 1(6): 612-619.

Keall, P., P. Poulsen and J. T. Booth (2019). "See, Think, and Act: Real-Time Adaptive Radiotherapy." Semin Radiat Oncol 29(3): 228-235.

Keall, P. J., G. S. Mageras, J. M. Balter, R. S. Emery, K. M. Forster, S. B. Jiang, J. M. Kapatoes, D. A. Low, M. J. Murphy, B. R. Murray, C. R. Ramsey, M. B. Van Herk, S. S. Vedam, J. W. Wong and E. Yorke (2006). "The management of respiratory motion in radiation oncology report of AAPM Task Group 76." Med Phys 33(10): 3874-3900.

Keall, P. J., A. Sawant, R. I. Berbeco, J. T. Booth, B. Cho, L. I. Cervino, E. Cirino, S. Dieterich, M. F. Fast, P. B. Greer, P. M. Af Rosenschold, P. J. Parikh, P. R. Poulsen, L. Santanam, G. W. Sherouse, J. Shi and S. Stathakis (2021). "AAPM Task Group 264: The safe clinical implementation of MLC tracking in radiotherapy." Medical Physics 48(5): E44-E64.

Kimura, T., T. Fujiwara, T. Kameoka, Y. Adachi and S. Kariya (2022). "The Current Role of Stereotactic Body Radiation Therapy (SBRT) in Hepatocellular Carcinoma (HCC)." Cancers (Basel) 14(18).

Langen, K. M. and D. T. L. Jones (2001). "Organ motion and its management." International Journal of Radiation Oncology Biology Physics 50(1): 265-278.

Li, R. J., J. H. Lewis, X. Jia, T. Y. Zhao, W. F. Liu, S. Wuenschel, J. Lamb, D. S. Yang, D. A. Low and S. B. Jiang (2011). "On a PCA-based lung motion model." Physics in Medicine and Biology 56(18): 6009-6030.

Lin, Y. Q., Z. J. Luo, W. Zhao and X. M. Li (2023). "Learning Deep Intensity Field for Extremely Sparse-View CBCT Reconstruction." Medical Image Computing and Computer Assisted Intervention, Miccai 2023, Pt X 14229: 13-23.

Liu, X., L. S. Geng, D. Huang, J. Cai and R. Yang (2024). "Deep learning-based target tracking with X-ray images for radiotherapy: a narrative review." Quant Imaging Med Surg 14(3): 2671-2692.

Lu, J., T. M. Guerrero, P. Munro, A. Jeung, P. C. Chi, P. Balter, X. R. Zhu, R. Mohan and T. Pan (2007). "Four-dimensional cone beam CT with adaptive gantry rotation and adaptive data sampling." Med Phys 34(9): 3520-3529.

Mildenhall, B., P. P. Srinivasan, M. Tancik, J. T. Barron, R. Ramamoorthi and R. Ng (2022). "NeRF: Representing Scenes as Neural Radiance Fields for View Synthesis." Communications of the Acm 65(1): 99-106.

Molaei, A., A. Aminimehr, A. Tavakoli, A. Kazerouni, B. Azad, R. Azad and D. Merhof (2023). "Implicit Neural Representation in Medical Imaging: A Comparative Survey." 2023 Ieee/Cvf International Conference on Computer Vision Workshops, Iccvw: 2373-2383.

Muller, T., A. Evans, C. Schied and A. Keller (2022). "Instant Neural Graphics Primitives with a Multiresolution Hash Encoding." Acm Transactions on Graphics 41(4).

Mylonas, A., J. Booth and D. T. Nguyen (2021). "A review of artificial intelligence applications for motion tracking in radiotherapy." J Med Imaging Radiat Oncol 65(5): 596-611.

Nakao, M., M. Nakamura and T. Matsuda (2022). "Image-to-Graph Convolutional Network for 2D/3D Deformable Model Registration of Low-Contrast Organs." IEEE Trans Med Imaging PP.

Poulsen, P. R., J. Jonassen, M. L. Schmidt and C. Jensen (2015). "Improved quality of intrafraction kilovoltage images by triggered readout of unexposed frames." Med Phys 42(11): 6549-6557.

Rashid, A., Z. Ahmad, M. A. Memon and A. S. M. Hashim (2021). "Volumetric Modulated Arc Therapy (VMAT): A modern radiotherapy technique - A single institutional experience." Pak J Med Sci 37(2): 355-361.

Reed, A. W., H. Kim, R. Anirudh, K. A. Mohan, K. Champley, J. G. Kang and S. Jayasuriya (2021). "Dynamic CT Reconstruction from Limited Views with Implicit Neural Representations and Parametric Motion Fields." 2021 Ieee/Cvf International Conference on Computer Vision (Iccv 2021): 2238-2248.
Roman, N. O., W. Shepherd, N. Mukhopadhyay, G. D. Hugo and E. Weiss (2012). "Interfractional Positional Variability of Fiducial Markers and Primary Tumors in Locally Advanced Non-Small-Cell Lung Cancer During Audiovisual Biofeedback Radiotherapy." International Journal of Radiation Oncology Biology Physics **83**(5): 1566-1572.
Sakata, Y., R. Hirai, K. Kobuna, A. Tanizawa and S. Mori (2020). "A machine learning-based real-time tumor tracking system for fluoroscopic gating of lung radiotherapy." Phys Med Biol **65**(8): 085014.
Segars, W. P., G. Sturgeon, S. Mendonca, J. Grimes and B. M. Tsui (2010). "4D XCAT phantom for multimodality imaging research." Med Phys **37**(9): 4902-4915.
Seppenwoolde, Y., H. Shirato, K. Kitamura, S. Shimizu, M. van Herk, J. V. Lebesque and K. Miyasaka (2002). "Precise and real-time measurement of 3D tumor motion in lung due to breathing and heartbeat, measured during radiotherapy." Int J Radiat Oncol Biol Phys **53**(4): 822-834.
Shao, H. C., Y. Li, J. Wang, S. Jiang and Y. Zhang (2023). "Real-time liver motion estimation via deep learning-based angle-agnostic X-ray imaging." Med Phys **50**(11): 6649-6662.
Shao, H. C., T. Mengke, T. Pan and Y. Zhang (2024). "Dynamic CBCT imaging using prior model-free spatiotemporal implicit neural representation (PMF-STINR)." Phys Med Biol **69**(11).
Shao, H. C., J. Wang, T. Bai, J. Chun, J. C. Park, S. Jiang and Y. Zhang (2022). "Real-time liver tumor localization via a single x-ray projection using deep graph neural network-assisted biomechanical modeling." Phys Med Biol **67**(11).
Shen, L., W. Zhao, D. Capaldi, J. Pauly and L. Xing (2022). "A geometry-informed deep learning framework for ultra-sparse 3D tomographic image reconstruction." Comput Biol Med **148**: 105710.
Shen, L. Y., J. Pauly and L. Xing (2022). "NeRP: Implicit Neural Representation Learning With Prior Embedding for Sparsely Sampled Image Reconstruction." Ieee Transactions on Neural Networks and Learning Systems.
Shen, L. Y., W. Zhao and L. Xing (2019). "Patient-specific reconstruction of volumetric computed tomography images from a single projection view via deep learning." Nature Biomedical Engineering **3**(11): 880-888.
Shieh, C. C., Y. Gonzalez, B. Li, X. Jia, S. Rit, C. Mory, M. Riblett, G. Hugo, Y. Zhang, Z. Jiang, X. Liu, L. Ren and P. Keall (2019). "SPARE: Sparse-view reconstruction challenge for 4D cone-beam CT from a 1-min scan." Med Phys **46**(9): 3799-3811.
Shirato, H., Y. Seppenwoolde, K. Kitamura, R. Onimura and S. Shimizu (2004). "Intrafractional tumor motion: lung and liver." Semin Radiat Oncol **14**(1): 10-18.
Tegunov, D. and P. Cramer (2019). "Real-time cryo-electron microscopy data preprocessing with Warp." Nat Methods **16**(11): 1146-1152.
Tewari, A., J. Thies, B. Mildenhall, P. Srinivasan, E. Tretschk, W. Yifan, C. Lassner, V. Sitzmann, R. Martin-Brualla, S. Lombardi, T. Simon, C. Theobalt, M. Niessner, J. T. Barron, G. Wetzstein, M. Zollhofer and V. Golyanik (2022). "Advances in Neural Rendering." Computer Graphics Forum **41**(2): 703-735.
Tong, F., M. Nakao, S. Q. Wu, M. Nakamura and T. Matsuda (2020). "X-ray2Shape: Reconstruction of 3D Liver Shape from a Single 2D Projection Image." 42nd Annual International Conferences of the Ieee Engineering in Medicine and Biology Society: Enabling Innovative Technologies for Global Healthcare Embc'20: 1608-1611.
van Aarle, W., W. J. Palenstijn, J. Cant, E. Janssens, F. Bleichrodt, A. Dabravolski, J. De Beenhouwer, K. Joost Batenburg and J. Sijbers (2016). "Fast and flexible X-ray tomography using the ASTRA toolbox." Opt Express **24**(22): 25129-25147.
Verellen, D., M. De Ridder, N. Linthout, K. Tournel, G. Soete and G. Storme (2007). "Innovations in image-guided radiotherapy." Nature Reviews Cancer **7**(12): 949-960.

Vergalasova, I. and J. Cai (2020). "A modern review of the uncertainties in volumetric imaging of respiratory-induced target motion in lung radiotherapy." Med Phys **47**(10): e988-e1008.

Wei, R., F. G. Zhou, B. Liu, X. Z. Bai, D. S. Fu, B. Liang and Q. W. Wu (2020). "Real-time tumor localization with single x-ray projection at arbitrary gantry angles using a convolutional neural network (CNN)." Physics in Medicine and Biology **65**(6): 065012.

Xu, Q., G. Hanna, J. Grimm, G. Kubicek, N. Pahlajani, S. Asbell, J. Fan, Y. Chen and T. LaCouture (2014). "Quantifying rigid and nonrigid motion of liver tumors during stereotactic body radiation therapy." Int J Radiat Oncol Biol Phys **90**(1): 94-101.

Yasue, K., H. Fuse, S. Oyama, K. Hanada, K. Shinoda, H. Ikoma, T. Fujisaki and Y. Tamaki (2022). "Quantitative analysis of the intra-beam respiratory motion with baseline drift for respiratory-gating lung stereotactic body radiation therapy." J Radiat Res **63**(1): 137-147.

Ying, X. D., H. Guo, K. Ma, J. Wu, Z. X. Weng and Y. F. Zheng (2019). "X2CT-GAN: Reconstructing CT from Biplanar X-Rays with Generative Adversarial Networks." 2019 Ieee/Cvf Conference on Computer Vision and Pattern Recognition (Cvpr 2019): 10611-10620.

Zha, R. Y., Y. H. Zhang and H. D. Li (2022). "NAF: Neural Attenuation Fields for Sparse-View CBCT Reconstruction." Medical Image Computing and Computer Assisted Intervention, Miccai 2022, Pt Vi **13436**: 442-452.

Zhang, C., L. Liu, J. Dai, X. Liu, W. He, Y. Chan, Y. Xie, F. Chi and X. Liang (2024). "XTransCT: ultra-fast volumetric CT reconstruction using two orthogonal x-ray projections fro image-guided radiation therapy via a transformer network." Phys. Med. Biol. **69**: 085010.

Zhang, Y., J. Ma, P. Iyengar, Y. Zhong and J. Wang (2017). "A new CT reconstruction technique using adaptive deformation recovery and intensity correction (ADRIC)." Med Phys **44**(6): 2223-2241.

Zhang, Y., H. C. Shao, T. Pan and T. Mengke (2023). "Dynamic cone-beam CT reconstruction using spatial and temporal implicit neural representation learning (STINR)." Phys Med Biol **68**(4).

Zhou, D., M. Nakamura, N. Mukumoto, M. Yoshimura and T. Mizowaki (2022). "Development of a deep learning-based patient-specific target contour prediction model for markerless tumor positioning." Med Phys **49**(3): 1382-1390.

Zhu, M., Q. Fu, B. Liu, M. Zhang, B. Li, X. Luo and F. Zhou (2024). "RT-SRTS: Angle-agnostic real-time simultaneous 3D reconstruction and tumor segmentation from single X-ray projection." Comput Biol Med **173**: 108390.

Zijp, L., J. J. Sonke and M. van Herk (2004). "Extraction of the respiratory signal from sequential thorax cone-beam X-ray images." ICCR: 507-509.